\documentclass{ws-procs10x7}
\usepackage{balance}
\usepackage{psfig}
\usepackage{pennames}

\newcolumntype{d}[1]{D{.}{.}{#1}}

\def\Journal#1#2#3#4{{\it #1} {\bf #2}, #3 (#4)}

\makeindex
\begin{document}

\title{STUDIES OF FRAGMENTATION AND \\
COLOUR RECONNECTION AT LEP}

\author{P. AZZURRI}

\address{Scuola Normale Superiore, Piazza dei Cavalieri 7, 56126 Pisa, Italy
\\E-mail: paolo.azzurri@sns.it}

\twocolumn[\maketitle\abstract{
Hadronic events at the Z pole have been investigated in search for 
Colour Reconnection effects and QCD coherence. 
Colour Reconnection effects are searched for in three-jet events
and the data results are compared to the predictions of different
Monte~Carlo models.
QCD colour coherence effects are tested through the multiplicity
distributions of hadrons with restricted momenta, and
the LEP1 $\Pep\Pem$ 
data is compared to HERA $\Pep\Pp$ data under the equivalence assumption 
of the $\Pep\Pem$ hemisphere with the current region of the Breit frame of reference.
}
]

\section{Colour Reconnection effects at the Z pole}
Colour Reconnection (CR) effects are higher-order and/or 
non perturbative QCD 
effects arising in the hadronisation of multi-parton systems.
There is a specific interest in these effects due to related uncertainties 
when performing precision measurements of invariant masses 
in multi-jet events, and in particular for
the measurement of the W mass, where possible CR effects 
in the $\PWp\PWm\rightarrow\Pq\Paq\Pq\Paq$ channel are
the main limitation on the overall $m_\PW$ precision. 

To assess CR effects it has been proposed\cite{Tc} 
to search for CR effects in
\Pq\Paq gg systems in hadronic Z decays,
benefiting greatly from larger statistics with respect to the W pair data.
In hadronic Z decays CR effects would enhance the probability to 
find a colour singlet gluon system isolated from the 
$\Pq\Paq$ state. Such events would show up as 
three-jet events with a zero charge gluon jet,
associated with a gap of particles around the gluon jet. 

\subsection{OPAL results}
The OPAL collaboration has selected gluon jets from Z decays
in events with two opposite tagged quark jets. A total of 439 gluon jets 
are selected with an expected purity of 82\% and an average energy of 40~GeV.
Studying the  distribution of charged particles at small
rapidities ($y\leq2$) around the gluon-jet axis, the expected 
multiplicity depletion due to CR effects is not observed, and 
the Ariadne\cite{ari} CR model was excluded at the level
of five standard deviations\cite{Oc1}.

In a subsequent publication by OPAL\cite{Oc2} 
lower energy gluon jets are selected in 
three-jet Z decays, where two jets are lifetime-tagged as 
heavy quark jets, and the third gluon jet has an average energy of 21~GeV. 
A gap in rapidity of both charged and neutral particles is ensured by
 requiring an upper bound on the minimum particle rapidity in the gluon jet
 ($y_{\rm min}<1.4$) and a lower bond on the maximum difference 
 between the rapidity of adjacent particles ($\Delta y_{\rm max}>1.3$). 
 The selection leads to a sample of 655 gluon jets with 
 an expected purity of 86\%.
 CR effects are expected to enhance the rates of those events for
 which the leading part of the gluon jet is neutral. 
 The enhancements predicted by the Ariadne\cite{ari} and GAL\cite{gal}
 CR models are not observed in the data, and retuning the parameters of 
 the two models to describe both the gluon jet data and the inclusive Z data,
 leads to a huge degradation of the $\chi^2$ values of the global fit, therefore
 excluding these two CR models. 
 For this analysis, the excess predicted by the Herwig CR model\cite{her} 
 is much less prominent, so that no definite conclusion could be obtained 
 on this model. 

\subsection{L3 results}
The L3 collaboration has also published studies of CR effects in  
three-jet hadronic Z decays\cite{Lc}. Events are selected with 
two lifetime-tagged quark jets and an anti-tagged gluon jet.
Observables based on angular separations of particles 
in the inter-jet regions are found to be very sensitive 
to colour singlet gluon productions and CR effects.
The data $\chi^2$ confidence level (CL) based on these observables
are then $\sim10^{-8}$ for the Ariadne CR model and $\sim10^{-6}$ for the 
GAL CR models, so that both these CR models are excluded with their
default settings. 
The same observables exclude also the Herwig\cite{her} model with a
$\sim10^{-9}$ CL including CR effects, but also exclude, with a 
$\sim10^{-8}$  CL, the same model with no CR effects,  
suggesting that Herwig can't simulate with sufficient 
precision the soft hadronisation effects important for CR studies.

\subsection{DELPHI results}
In recent preliminary studies from DELPHI\cite{Dc}, three-jet events are 
selected from Z decays, and the quark or gluon jets are singled out by energy 
ordering. The leading system of both kind of jets are selected by demanding 
a rapidity gap of charged particles for $\Delta y \leq 1.5$. In this way about 50,000
gluon jets and 50,000 quark jets are selected. 
Studying the total charge of the leading system, a higher rate of neutral systems 
is found in the gluon jets data with respect to the predictions of string models,
while  no such enhancement is found in the quark jets sample. 
The excess of leading neutral gluon jets is measured to be roughly 10\% with
a significance of three standard deviations, and could be due to CR effects or 
colour-octet neutralisation of the gluon field (glueballs).

\subsection{ALEPH results}
In a new paper by the ALEPH collaboration\cite{Ac}
Z decays to three jets are also used to test CR models. 
The main fragmentation parameters of the CR models 
considered\cite{ari,gal,her} have been re-tuned with
fits to the hadronic Z global event shape and charged particle 
momentum distributions, and without degrading the fits 
$\chi^2$ values with respect to the models with no CR.

To select three-jet events the Durham clustering scheme is applied
to hadronic Z decays with a resolution parameter $y_{\rm cut}=0.02$,
and only events which cluster into exactly three-jets are selected.
The events are also required to be planar, with no isolated and energetic photon. 
The three jets are required to be well-contained in the detector acceptance
and separated by inter-jet angles larger than 40$^\circ$.
Jets are energy-ordered and the softest jet (jet 3) is assigned to the gluon jet,
with an expected purity of 70\%.

\begin{figure}[hbt]
\centerline{\psfig{file=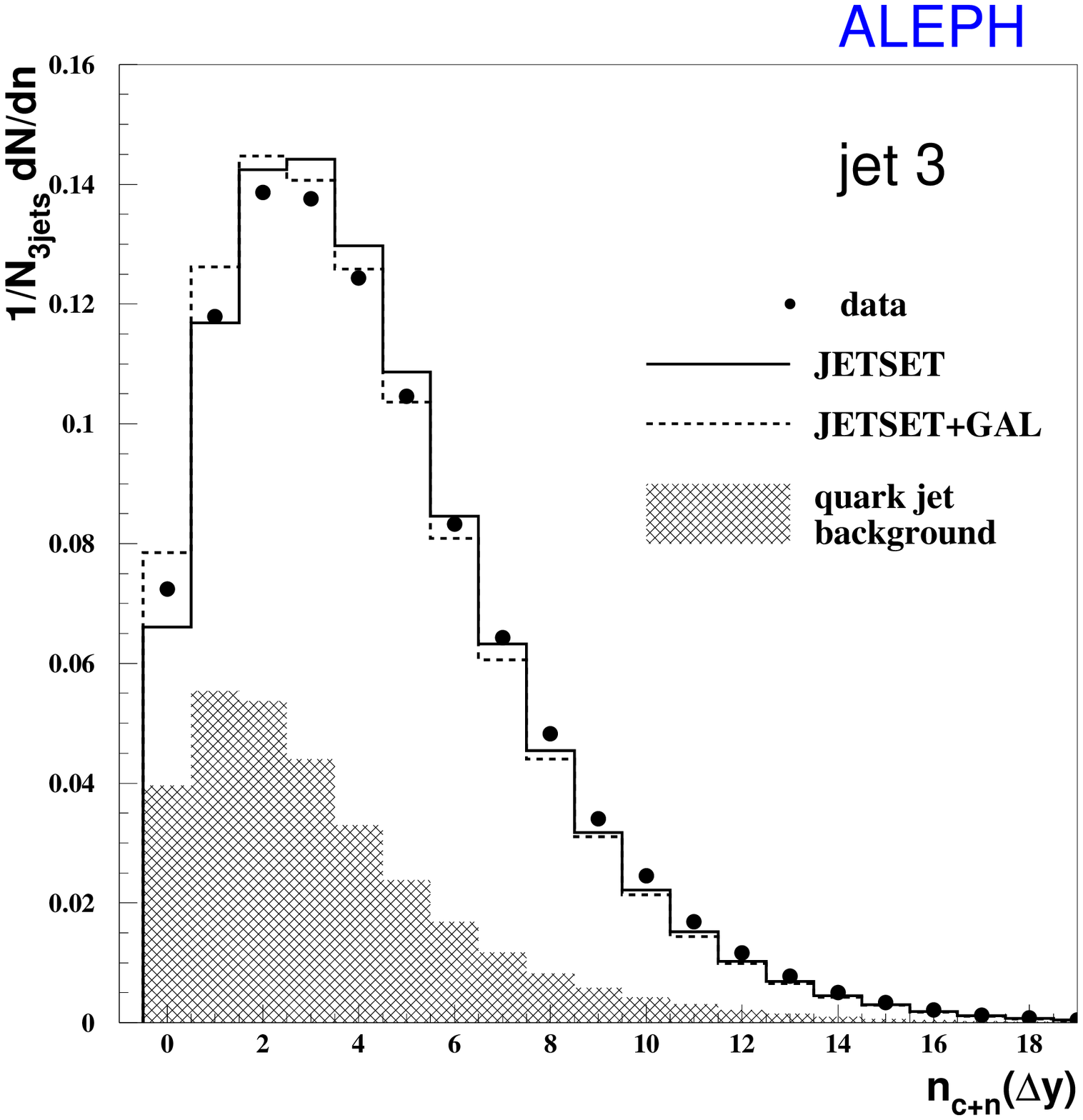,width=4cm}
\hspace{-0.4cm}
\psfig{file=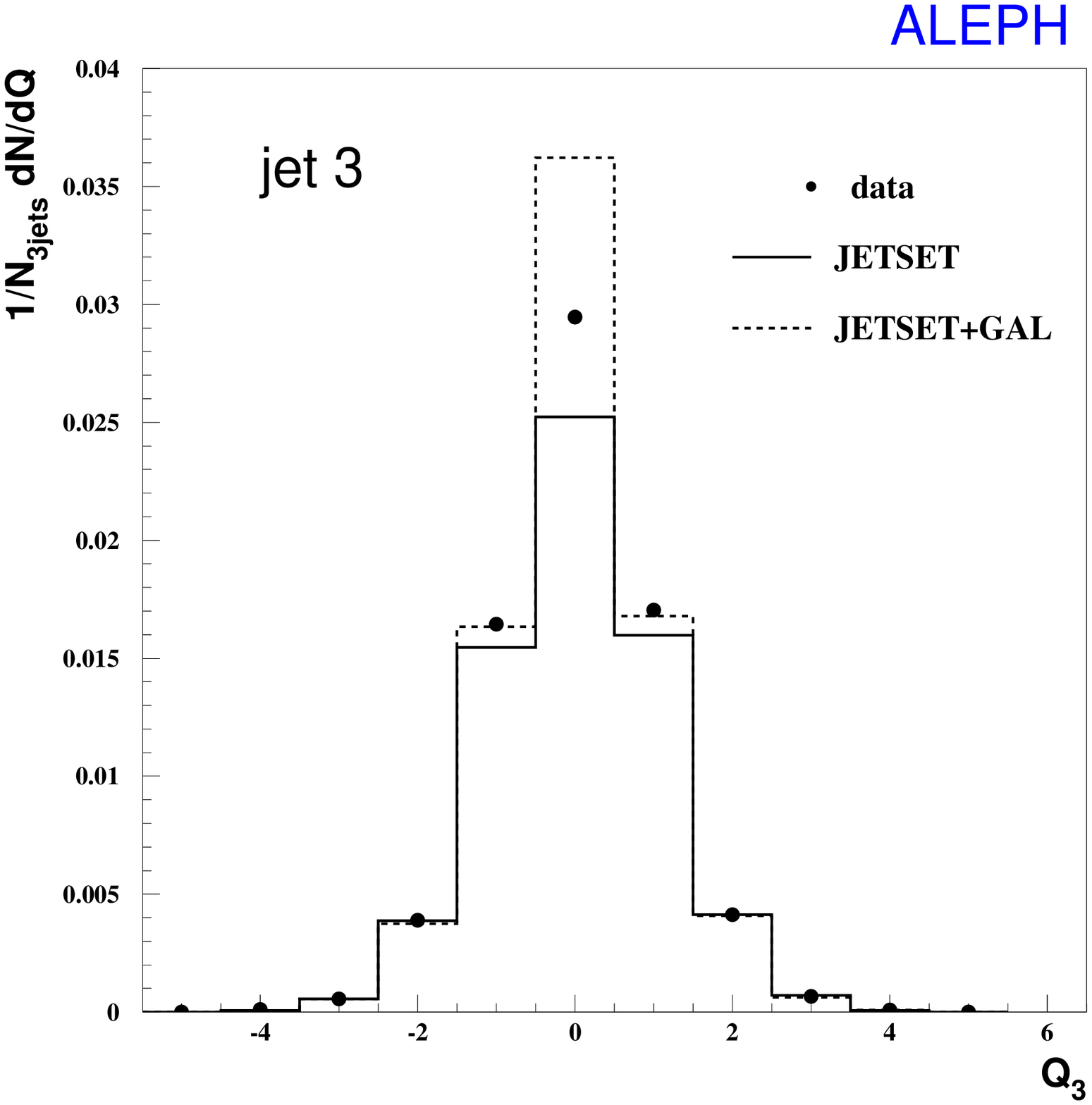,width=3.8cm}}
\caption{Number of charged and neutral particles in the $0<y<1.5$ interval of jet 3,
compared to the prediction of a fragmentation model 
with and without GAL CR effects~\protect\cite{gal} (left). 
Charge distribution of jet 3 with a rapidity gap in the $0<y<1.5$ interval (right).
\label{fig1}
}
\end{figure}

The particle multiplicity within the central rapidity interval $0<y<1.5$ of jet 3
is shown in figure~\ref{fig1}(left). 
Focusing on the first bin with no particles in the rapidity interval (rapidity gap)
it can be seen that there is an excess of data events with respect to the predictions 
of the fragmentation model with no CR effects, but the data excess is  smaller than the  
predictions of the same model with CR effects\cite{gal}.
The differences of the data yields and the models are enhanced 
by requiring zero total charge for jet 3, as visible in figure~\ref{fig1}(right).

\begin{figure}[hbt]
\centerline{\psfig{file=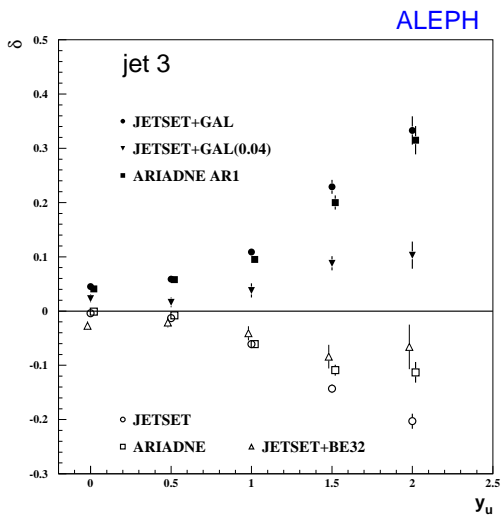,width=7cm}}
\caption{Relative model-data differences in the rate of neutral gluon jets
as a function of the required rapidity gap. \label{fig2}
}
\end{figure}

The differences $\delta$=(model-data)/data
of the rate of neutral gluon jets, as a function of 
the required rapidity gap is shown in figure~\ref{fig2}
for different models. It can be seen that at large rapidity 
gaps the data is in disagreement both with models with
or without CR effects included. 
To fit the data rates with the Ariadne and GAL CR models, 
it would be necessary to decrease their CR 
strength parameters  roughly by a factor five, 
from 0.1 to 0.02, in disagreement with the $1/N^2_c\simeq 1/9$ 
prescription for the reconnection probability.

\begin{figure}[hbt]
\centerline{\psfig{file=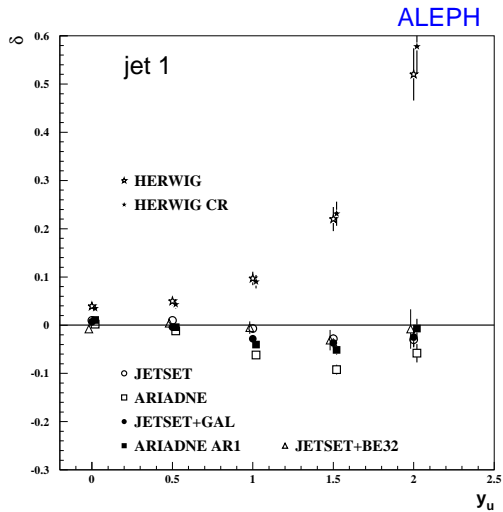,width=7cm}
}
\caption{Relative model-data differences in the rate of neutral quark jets
as a function of the required rapidity gap. \label{fig3}
}
\end{figure}

Figure~\ref{fig3} shows the $\delta$ values for the leading quark jet.
In this case the agreement of data with models with or without CR effects
is good, except for the Herwig model that in both cases fails to reproduce 
the particle rapidity distributions, over a wide rapidity range,
and is therefore not suited for these studies.

\section{QCD Coherence and Correlations at the Z pole}
Recent analytical perturbative QCD calculations,
in conjunction with Local Parton-Hadron Duality (LPHD),
suggest that, while the general multiplicity distribution of 
partons in a jet is broader than a Poisson distribution, due to positive correlations 
in the gluon emissions, for gluons produced with limited momenta transverse 
to the primary parton, the emissions are independent due to colour coherence,
and their multiplicity distribution becomes Poissonian\cite{Tq}. 

To verify this perturbative prediction and to what extent it is affected by the 
hadronisation, it has been proposed to measure factorial moments $F_q$
and cumulants $K_q$, defined as
\begin{eqnarray*}
F_q= \langle n (n-1) \hdots (n-q-1)\rangle / \langle n \rangle^q  \\
K_2 = F_2-1 \hspace{.5cm}
K_3=F_3-3F_2+2
\\ K_4=F_4-4F_3-3F_2^2+12F_2-6 
\end{eqnarray*}
where $n$ is the number of particles in some phase-space region, 
and the angle 
brackets denote the average over the observed events. 
The $K_q$ moments are by construction the measure of 
the $q$-particle correlations.

Restricting the phase-space of the multiplicity measurements 
cylindrically in transverse momentum ($p_T<p_T^{\rm cut}$)
the Poisson limit is expected to be reached as $p_T^{\rm cut}$ decreases,
with $F_q\simeq 1$ and $K_q\simeq 0$. 
The same Poisson limit is not expected to be 
reached when limiting the phase-space 
spherically in absolute momentum ($p<p^{\rm cut}$)\cite{Tq}.

\begin{figure}[hbt]
\centerline{\psfig{file=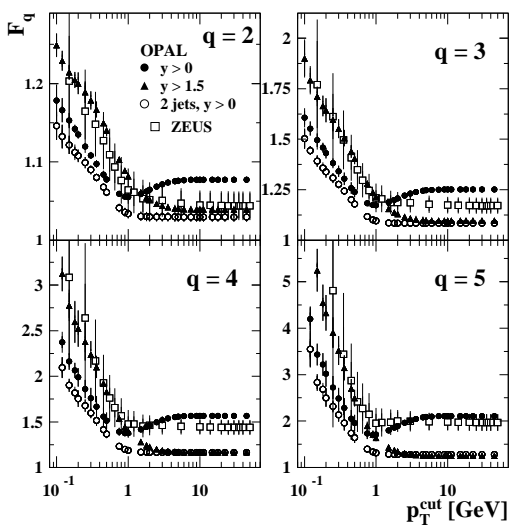,width=7.5cm}}
\caption{Factorial moments $F_q$ of charged particles with limited transverse momenta,
as a function of $p_T^{\rm cut}$, compared to those measured in high thrust two jet events, 
in a restricted rapidity window, and in HERA $\Pep\Pp$ data from ZEUS~\protect\cite{Zq}.
 \label{fig4}}
\end{figure}

The OPAL collaboration has recently published results on the measurements
of these proposed correlation variables\cite{Oq}. 
Results for cylindrically cut $F_q$ moments are shown in figure~\ref{fig4}.
Decreasing $p^{\rm cut}$ the moments decrease to a minimum at the 
common value $p^{\rm cut}\simeq 1$~GeV, suggesting that the strong 
hadronisation effects mask the validity of the perturbative calculations at the scale of 1~GeV. 

Results on cumulants reveal similar structures and show evidence for the presence of two- 
and three-particle correlations, while four-particle correlations $K_4$ are compatible with zero,
within errors.

Results on $F_q$ moments measured by ZEUS\cite{Zq} in $\Pep\Pp$ data are also 
shown in figure~\ref{fig4}  but do not reveal the same minimum that could signal the 
border between perturbative and non-perturbative dynamics. 

To understand the differences, further cuts have been
applied to the LEP $\Pep\Pem$ data to mimic the HERA $\Pep\Pp$ 
experimental conditions, including (i) a cut $y>1.5$ to exclude the 
central rapidity region and (ii)  a cut on the event thrust $T>0.96$ to select
a pure sample of two-jet events.

With the additional cuts, also shown in figure~\ref{fig4},
the qualitative behaviour of the $\Pep\Pem$ 
and  $\Pep\Pp$ measurements looks more similar, 
but the compatibility of the two measurements is still problematic.
These results suggest that for soft particle production, 
the assumed equivalence 
of a single event hemisphere in  $\Pep\Pem$ events with 
the current region in the $\Pep\Pp$  Breit frame may be misleading.

\end{document}